\documentclass{PoS}

\newcommand{\NP}[1]{ Nucl.\ Phys.\ {#1}}

\newcommand{\PR}[1]{Phys.\ Rev.\ { #1}}
\newcommand{\PRL}[1]{ Phys.\ Rev.\ Lett.\ { #1}}

\newcommand{\be}{\begin{equation}}
\newcommand{\ee}{\end{equation}}
\newcommand{\ba}{\begin{eqnarray}}
\newcommand{\ea}{\end{eqnarray}}

\newcommand{\ima}{{\mbox{Im}\,}}

\PoS{PoS(HEP2005)149}

\title{Strong unitary violations in the extra dimensional SM and the equivalence
between Goldstone and longitudinal gauge bosons.}

\ShortTitle{Strong unitary violations in the extra dimensional SM}

\author{\speaker{Jos\'e R. Pelaez}\thanks{Work partially supported by an INFN-CICYT collaboration project.}\\

        Dept. F\'isica Te\'orica II. Universidad Complutense, Spain\\

        E-mail: \email{jrpelaez@fis.ucm.es}}

\author{S. De Curtis and D. Dominici\\

        INFN Sezione di Firenze and Dip. di Fisica Universit\'a degli Studi
, Firenze.\\

       E-mail: \email{decurtis@fi.infn.it}, \email{dominici@fi.infn.it}
}

\abstract{Tree level unitarity violations of extra dimensional extensions of the Standard Model may 
become much stronger when the scalar sector is included in the bulk.
This effect occurs when the couplings are not suppressed for larger Kaluza-Klein
levels, and could have relevant consequences for the phenomenology of the next generation of colliders. We briefly review our formalism 
to obtain more stringent unitarity bounds when KK modes are present, as well
as the generalization to extra dimensions 
of the Equivalence Theorem between Goldstone
 bosons and longitudinal gauge bosons.}

\FullConference{International Europhysics Conference on High Energy Physics\\

                 July 21st - 27th 2005\\

                 Lisboa, Portugal}

\begin{document}

\section{Introduction.}

Violations of tree level unitarity are an indication
of a perturbation theory breakdown and
that the interaction has become strong.
As it is well known, this happens in the Standard Model (SM)
for Higgs masses of the order of 1 TeV. 
 Let us now recall how this bound was obtained \cite{Lee:1977eg}. First, we need
the definition of partial
waves of angular momentum $J$ for two-body scattering,
and the familiar unitarity condition they satisfy
in the elastic regime, when only one state  $\alpha$ is available:
\begin{eqnarray*}
t_{\alpha\beta}^{J}(s)
\equiv\frac{1}{32\pi}\int^1_{-1}\,d(\cos\theta) 
T_{\alpha\beta}(s,t,u)
P_J(\cos\theta),\quad\stackrel{\hbox{\footnotesize elastic}}{\longrightarrow}\quad
\ima t^J_{\alpha\alpha}  = \sigma_\alpha \,\vert
 \,t^J_{\alpha\alpha} \vert ^2, \quad\sigma_\alpha=2q_\alpha/\sqrt{s},
\nonumber
\end{eqnarray*}
 $q_\alpha$ being the CM momenta.
Note that in the $s\rightarrow\infty$ limit, $\sigma_\alpha\rightarrow1$ 
and unitarity implies $|t_{\alpha\alpha}|^J\leq1$.
When more states, $\beta$, are accessible, we form a matrix 
$T^J_{\alpha\beta}$ and unitarity requires
its eigenvalues to be smaller than one.
For example, 
in the SM we have tree level couplings between
$\alpha=HH/\sqrt{2}$, 
$\beta=Z_{L}Z_{L}/\sqrt{2}$, $\gamma=W_{L}^+W_{L}^-$.
Thus, in the $s\rightarrow\infty$ 
limit, $T^{J=0}_{\alpha\beta}$ reads
  \begin{eqnarray*}
T^{J=0}=\frac{G_F m_h^2}{4\pi\sqrt{2}}\left(
   \begin{array}{ccc}
1&1/\sqrt{8}&1/\sqrt{8}\\
1/\sqrt{8}&3/4&1/4\\
1/\sqrt{8}&1/4&3/4
   \end{array}
\right)\;
\stackrel{\hbox{\footnotesize Eigenvalues}}{ \longrightarrow}\;
\frac{G_F m_h^2}{8\pi\sqrt{2}}(3,1,1)\le1
\;
\stackrel{\hbox{\footnotesize Unitarity}}{ \Longrightarrow}\;
m_h^2\le\frac{8\pi\sqrt{2}}{3 G_F}.
\end{eqnarray*}

\section{Unitarity conditions including Kaluza Klein excitations.}

In a recent work \cite{DeCurtis:2003zt},
 we have shown that unitarity bounds can become more stringent
in extra dimensional extensions of the SM.
In fact, {\it any additional 
Kaluza-Klein (KK) state}, $\beta$, {\it coupled to the SM states above}, $\alpha$,
{\it always provide additional contributions to
the unitarity bound}, i.e.
\be
\hbox{Unit}_{\alpha\rightarrow\alpha}\equiv
\sigma_\alpha\vert t^J_{\alpha\alpha}\vert+
\frac{1}{\vert t^J_{\alpha\alpha}\vert}\sum_{\beta\neq\alpha}
\sigma_\beta \,\vert\, t^J_{\alpha\beta} \vert ^2 \le 1.
\label{newunitbound}
\ee
Thus, {\it tree level unitarity violations occur for smaller $m_h$ and lower
$s$ than in the SM.}
%, as long as the new amplitudes $t^J_{\alpha\beta}$ 
%are not sufficiently suppressed}. 
Of course,  in models where KK states do not couple
to Higgs or longitudinal gauge bosons, or where their tree level couplings
are sufficiently suppressed, Eq.(\ref{newunitbound}) still holds,
but the new terms may be negligible.
However we have shown that this 
effect can be rather large when Higgs fields extend in the bulk
and their KK modes have the same couplings as the zero mode.
This is the case, for instance, of universal extra dimensions \cite{Barbieri:2000vh}.
Here we will briefly review how, with compactification scales of the order of
$1/R\simeq0.5-3\,$TeV, this effect could lead to a strongly interacting regime
within the reach of the LHC. In order to do so, we first review 
a useful tool to calculate longitudinal gauge boson amplitudes, and its extension
to extra dimensions.

\section{The Equivalence Theorem in extra dimensions.}

In renormalizable $R_\xi$ gauges, the gauge fixing function $f_a=\partial_\mu V^{\mu a}-\xi M G^{V a}$,  which acts like a delta function in field space,
implies $p_\mu V^{\mu a}=\xi M G^{V a}$. 
Since $\epsilon_{L\mu}=p_\mu/M+O(M/\sqrt{s})$,
%this means that longitudinal gauge bosons 
then $V_L\equiv\epsilon_{L\mu}V^{\mu}\simeq G^V$ and,
up to $O(M/\sqrt{s})$, 
the Equivalence Theorem \cite{ET} allows us to calculate amplitudes for
longitudinal gauge bosons
by replacing them by their associated Goldstone bosons.

We have used an extra dimensional generalization \cite{DeCurtis:2002nd}  
with 5D gauge bosons and Higgs fields in the bulk, 
on the brane, or both. The novelty is that mass matrices are different
now for $V_L$ and $G^V$ and are non diagonal,
mixing the Goldstone boson KK modes $G^V_{(n)}$ with the $V^{5}_{(n)}$
(the latter give masses to the KK $V_L$ 
in the unbroken case \cite{SekharChivukula:2001hz}).
If the Higgs is on the brane there is also KK level mixing.
We thus had to perform a rotation
to find the gauge boson mass eigenstates
and a different one to diagonalize the Goldstone
 boson ``mass'' matrix $M_G^2$.
It was not trivial to find that the mixing term from the
gauge fixing function, proportional to $M_G$ and not to $M_G^2$,
becomes diagonal under these simultaneous rotations. This implies that there is a 
complicated combination of Goldstone bosons and $V^5_{(n)}$ KK modes,
that is a ``mass'' eigenstate, which is related (``eaten'')
by their associated longitudinal gauge boson mass eigenstate,
so that,
as usual,
\begin{equation}
  T(\hat V^{\mu }_{L\,(m)},\hat V^{\mu }_{L\,(n)},\ldots)
\simeq  C^{(m)} C^{(n)}... T(\hat G^V_{(m)},\hat G^V_{(n)},\ldots) +O(M_k/E)
\end{equation}
$M_k$ being the largest
$m_{V(m)}$ mass, and $C^ {(i)}=1+$ 
``electroweak corrections''\cite{ET}.
Since the $\hat G^V$ are scalars, in the Landau gauge
the calculations are considerably simplified.
The $a_{(n)}$ combinations of $G^V$ and $V^5$ not ``eaten''
by the $V_L$, remain
massive with $m_{a(n)}^2=m_{h(0)}^2+n^2/R^2$.

\section{An example of strong unitarity violations with scalars in the bulk.}

Let us consider a minimal 5D extension
of the SM  compactified on the segment $S^1/Z_2$, of length $\pi
R$, in which the $SU(2)_L$ and $U(1)_Y$ gauge and Higgs fields 
propagate in the bulk. It is straightforward 
to write
the Lagrangian and tree level partial waves \cite{DeCurtis:2003zt}
of the state $\alpha=W_{L\,(0)}^+W_{L\,(0)}^-$ scattering into 
itself,
as well as into 
$\beta=h_{(0)} h_{(0)}/\sqrt{2}$,
$Z_{L\,(0)}Z_{L\,(0)}/\sqrt{2}$, $h_{(n)} h_{(n)}/\sqrt{2}$,
$a_{(n)}^3 a_{(n)}^3/\sqrt{2}$ and $a_{(n)}^+ a_{(n)}^-$, 
whose interactions are not
suppressed by any power of $R$ or EW couplings.

In Figure 1 we plot the left hand side of Eq.(\ref{newunitbound}), 
for $1/R=0.5\,$TeV and 3 TeV, and we see that tree level unitarity violations
occur for smaller values of $m_{h(0)}$ and $\sqrt{s}$ than in the SM.

\begin{figure}[htbp]
  \centering
  \includegraphics[width=.6\textwidth]{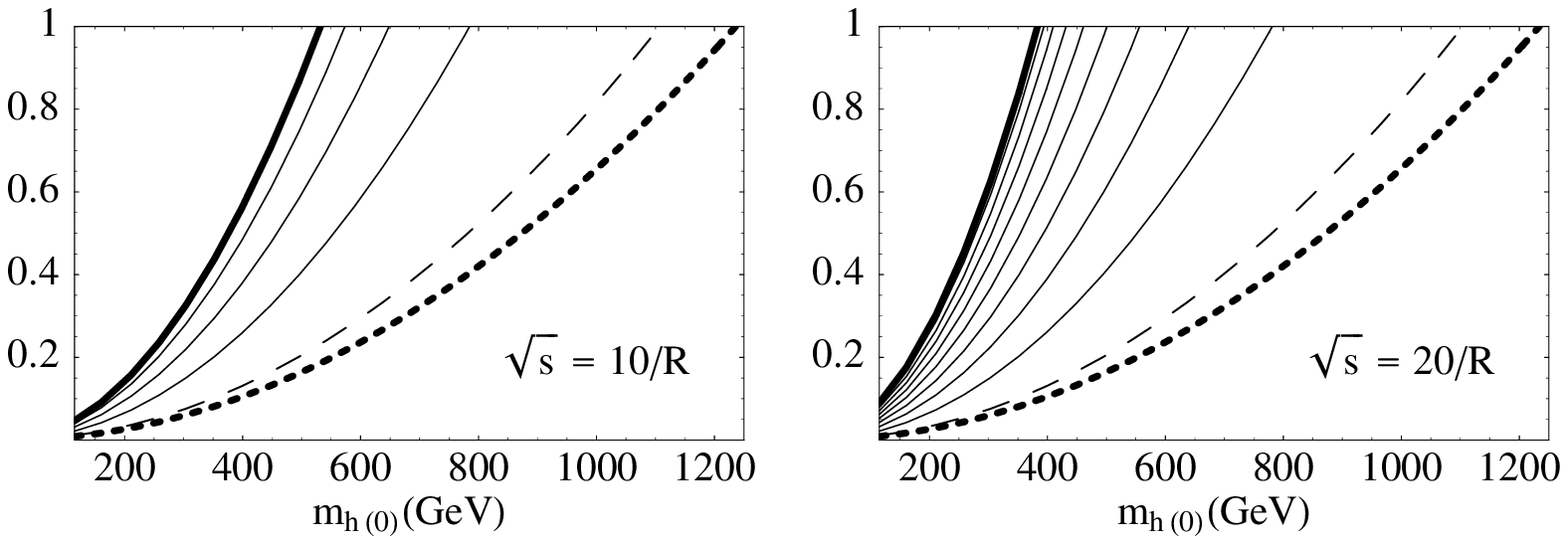}
  \includegraphics[width=1.\textwidth]{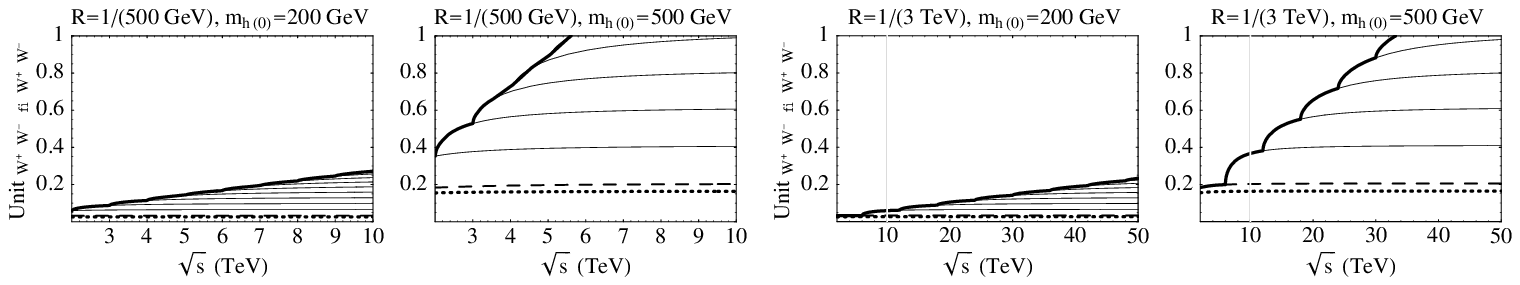}
  \caption{The unitarity bound,
Eq.(\protect\ref{newunitbound}) is violated above 1.
The dotted line is the SM bound using
only $W_ {L\,(0)}^+W_{L\,(0)}^-$ elastic scattering, 
whereas the dashed line 
includes also the $h_{(0)}h_{(0)}$,
$Z_ {L\,(0)}Z_{L\,(0)}$.
The continuous lines are the contributions of
the KK excitations, which, in total, yield the thick continuous line.
}
  \label{fig:2}
\end{figure}

In Figure 2 we plot in the $(\sqrt{s},m_{h{(0)}})$ plane the regions
where tree level unitarity is violated (white). Since we have only used
two-body states to derive Eq.(\ref{newunitbound}), and other, many body,
states should also contribute positively to the sum, we have also
plotted in gray the area where
Eq.(\ref{newunitbound}) is larger than 0.5. 
{\it In these two regions the electroweak symmetry breaking sector will most likely
become strongly interacting}. 
Finally, we can also see how the more dimensions are added ($D=6,7...$),
the stronger tree level unitarity violations become.
Finally, let us remark that {\it this effect
 could be accessible at the next generation of colliders, like LHC}.

\begin{figure}[htbp]
  \centering
  \includegraphics[width=\textwidth]{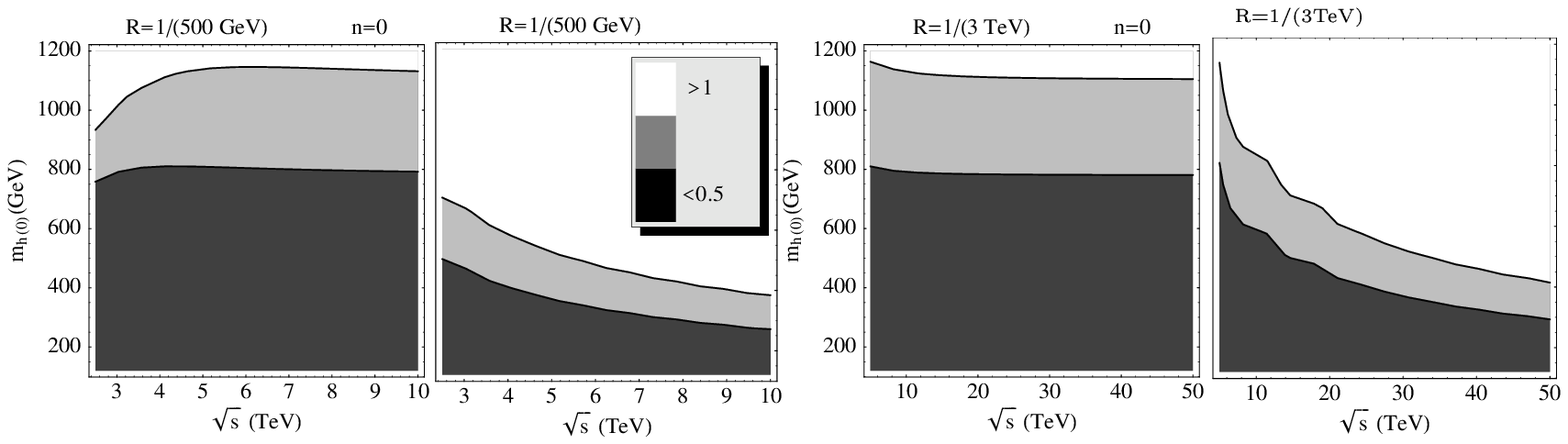}
  \includegraphics[width=.85\textwidth]{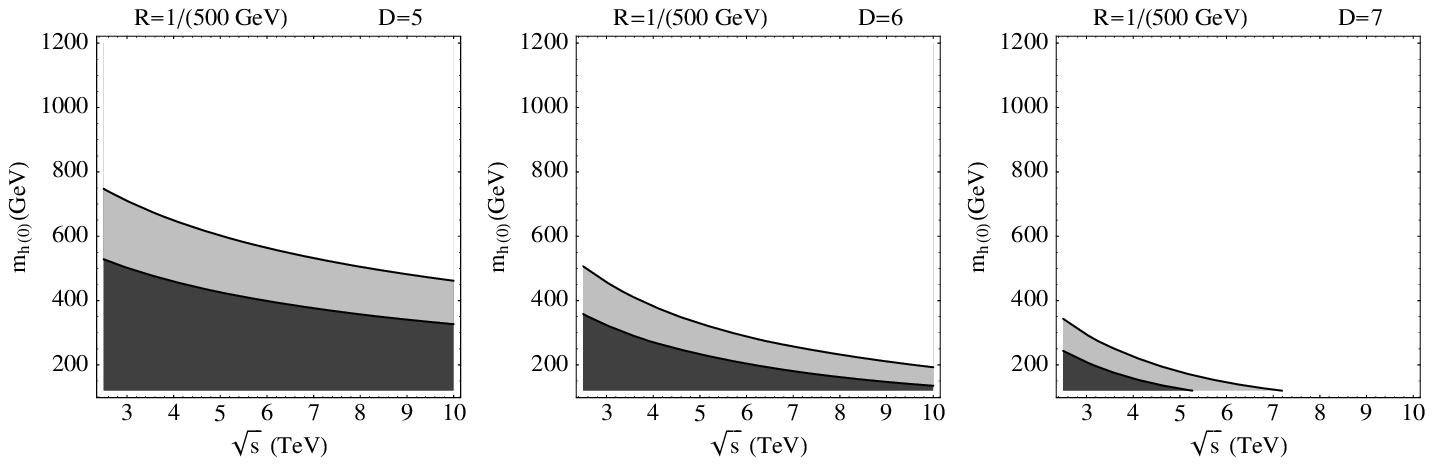}
  \caption{Tree level unitarity violation (white) and 
possibly strongly interacting regions (gray) for different values of
$R$, $\sqrt{s}$, $m_{h(0)}$ and dimensions $D$,
for models with unsuppressed KK scalar couplings.}
  \label{fig:4}
\end{figure}

\end{document}